\documentclass[proof]{pasj01}
\Received{$\langle$reception date$\rangle$}
\Accepted{$\langle$acception date$\rangle$}
\Published{$\langle$publication date$\rangle$}

\usepackage{mathpazo}
\usepackage[T1]{fontenc}

\begin{document}

\title{ {\LARGE LETTER} \\
First Detection of HC$_{5}$$^{15}$N in the Interstellar Medium}
\author{Kotomi \textsc{taniguchi}\altaffilmark{1,2}}
\altaffiltext{1}{Department of Astronomical Science, School of Physical Science, SOKENDAI (The Graduate University for Advanced Studies), Osawa, Mitaka, Tokyo 181-8588, Japan}
\altaffiltext{2}{Nobeyama Radio Observatory, National Astronomical Observatory of Japan, Minamimaki, Minamisaku, Nagano 384-1305, Japan}
\email{kotomi.taniguchi@nao.ac.jp}

\author{Masao \textsc{saito} \altaffilmark{1,2}}
\email{masao.saito@nao.ac.jp}

\maketitle

\begin{abstract}
We report the first detection of HC$_{5}$$^{15}$N with the $J=9-8$ rotational line from the cyanopolyyne peak in Taurus Molecular Cloud-1 (TMC-1 CP) using the 45-m radio telescope of the Nobeyama Radio Observatory.
The column density of HC$_{5}$$^{15}$N is derived to be ($1.9 \pm 0.5$)$\times 10^{11}$ cm$^{-2}$ ($1\sigma$).
We apply the double isotope method to derive the $^{14}$N/$^{15}$N ratios of HC$_{5}$N and HC$_{3}$N in TMC-1 CP.
The $^{14}$N/$^{15}$N ratios are calculated to be $344 \pm 53$ and $257 \pm 54$ for HC$_{5}$N and HC$_{3}$N, respectively.
The $^{14}$N/$^{15}$N ratio of HC$_{5}$N is lower than the elemental ratio in the local interstellar medium ($\sim 440$) and slightly higher than that of HC$_{3}$N in TMC-1 CP.
Since HC$_{3}$N is formed via the neutral-neutral reaction between C$_{2}$H$_{2}$ and CN, the slightly higher $^{14}$N/$^{15}$N ratio of HC$_{5}$N may support our previous suggestions that the main formation mechanism of HC$_{5}$N is the ion-molecule reactions between hydrocarbon ions (C$_{5}$H$_{n}^{+}$) and nitrogen atoms.
\end{abstract}

\section{Introduction} \label{sec:intro}

Carbon-chain molecules are unique species in the interstellar medium, and the studies about their chemical mechanisms have been progressed mainly by radio astronomical observations.
Survey observations showed that carbon-chain molecules are good chemical evolutional tracers \citep{1992ApJ...392..551S, 2009ApJ...699..585H, 2017ApJS..228...12T}.
Carbon-chain molecules such as CCS are abundant in young low-mass dark clouds and decrease in evolved star-forming cores.
The tendency is explained by the chemical characteristics of carbon-chain species; they are formed from carbon cations (C$^{+}$) or carbon atoms (C) \citep{1992ApJ...392..551S}, destroyed by the reactions with H$^{+}$, He$^{+}$, or O, and depleted onto dust grains \citep{2013ChRv...113...8981}.
However, since carbon-chain species have unsaturated bonds, they are unstable, and hence it is difficult to derive their main formation mechanisms from laboratory experiments.
Consequently, the chemical network model calculations about carbon-chain molecules have large uncertainties despite many attempts.

Recent development of the radio astronomical equipment allows us to detect weak lines, including isotopologues of carbon-chain molecules, within reasonable time.
In order to investigate main formation mechanisms of carbon-chain molecules using their $^{13}$C isotopic fractionation (the differences in abundance among the $^{13}$C isotopologues), various observations deriving $^{13}$C isotopic fractionation have been carried out in HC$_{3}$N \citep{1998A&A...329.1156T}, HC$_{5}$N \citep{2016ApJ...817..147T}, CCH \citep{2010A&A...512A...31S}, CCS \citep{2007ApJ...663...1174}, C$_{3}$S, and C$_{4}$H \citep{2013JPCA...117...9831}  toward the cyanopolyyne peak in Taurus Molecular Cloud-1 (TMC-1 CP; $d = 140$ pc), and in {\it {cyclic}}-C$_{3}$H$_{2}$ \citep{2015ApJ...807...66Y} toward the low-mass star-forming region L1527 ($d = 140$ pc). 

The main formation pathway of HC$_{3}$N in TMC-1 CP was suggested as the neutral-neutral reaction between C$_{2}$H$_{2}$ and CN, from the abundance ratios of $[$H$^{13}$CCCN$]: [$HC$^{13}$CCN$]: [$HCC$^{13}$CN$]$ = $1.0 : 1.0 : 1.4$ \citep{1998A&A...329.1156T}.
The observed abundance ratios can be explained by the reaction of C$_{2}$H$_{2}$ + CN, because C$_{2}$H$_{2}$ has two equivalent carbon atoms and $^{13}$C tends to concentrate in CN via the exothermic reaction between $^{13}$C$^{+}$ and CN \citep{1991ApJ...379..267K}.
\citet{2016ApJ...830..106T} also carried out observations deriving $^{13}$C isotopic fractionation of HC$_{3}$N toward the low-mass star-forming region L1527 and the high-mass star-forming region containing a hot core G28.28-0.36 ($d=3$ kpc). 
They suggested that the main formation pathways of HC$_{3}$N in the both star-forming regions are the same one as that in TMC-1 CP (C$_{2}$H$_{2}$ + CN).

On the other hand, \citet{2016ApJ...817..147T} proposed that the main formation mechanism of HC$_{5}$N in TMC-1 CP is the ion-molecule reactions between hydrocarbon ions (C$_{5}$H$_{n}^{+}$; $n=3-5$) and nitrogen atoms followed by the electron recombination reactions, based on the observational results showing the abundance ratios of $[$H$^{13}$CCCCCN$]: [$HC$^{13}$CCCCN$]: [$HCC$^{13}$CCCN$]: [$HCCC$^{13}$CCN$]: [$HCCCC$^{13}$CN$]$ = $1.00:0.97:1.03:1.05:1.16$ ($\pm 0.19$) (1$\sigma$).
In the proposed ion-molecule reactions, all carbon atoms in HC$_{5}$N originate from the hydrocarbon ions. 
Such large hydrocarbon ions are produced through various processes, and there is no reason that $^{13}$C is concentrated in a particular carbon atom in large hydrocarbon ions. 
In other words, there should be no clear $^{13}$C isotopic fractionation.
One difficulty using the $^{13}$C isotopic fractionation method is that the differences in the $^{12}$C/$^{13}$C ratio of each isotopologue are small, and we need long integration time to obtain spectra with sufficient signal-to-noise ratios.

In the present letter, we report the first detection of HC$_{5}$$^{15}$N from TMC-1 CP.
We derive its column density and the $^{14}$N/$^{15}$N ratio of HC$_{5}$N.
From the $^{14}$N/$^{15}$N ratio, we suggest the ion-molecule reactions as the main formation mechanism of HC$_{5}$N in TMC-1 CP also suggested from our previous work of the $^{13}$C isotopic fractionation of HC$_{5}$N.

\section{Observations} \label{sec:obs}

We carried out observations of HC$_{5}$$^{15}$N ($J=9-8$; 23.37544 GHz \citep{2005JChPh.122s4319S}) with the Nobeyama 45-m radio telescope during 2014 December and 2015 January (2014-2015 season)\footnote{\citet{2016ApJ...817..147T} described the observation date as 2014 March, April (2013-2014 season), December and 2015 January (2014-2015 season). In 2014 March and April, we carried out observations only in the 42 GHz band using the Z45 receiver. The observations in the 23 GHz band using the H22 receiver were conducted only in 2014 December and 2015 January; the data presented here were taken simultaneously with the normal species and the five $^{13}$C isotopologues of HC$_{5}$N in the 23 GHz band in 2014 December and 2015 January.}.
The observed position was ($\alpha_{2000}$, $\delta_{2000}$) = (04$^{\rm h}$41$^{\rm m}$42\fs49, 25\arcdeg41\arcmin27\farcs0) for TMC-1 CP.
The off-source position was set to be +30' away in the right ascension.
We checked the telescope pointing every 1.5 hr by observing the SiO maser line ($J=1-0$) from NML Tau.
The pointing error was less than 3".

We used the H22 receiver, which enables us to obtain dual polarization data simultaneously.
The H22 receiver is a single sideband (USB) receiver with its gain above 25 dB.
The beam size and the main beam efficiency ($\eta_{B}$) were 72" and 0.8, respectively. 
The system temperatures were from 90 to 110 K, depending on the weather conditions and elevations.
We used the SAM45 FX-type digital correlator in frequency setups whose bandwidths and frequency resolutions were 63 MHz and 15.26 kHz, respectively.
The frequency resolution corresponds to the velocity resolution of 0.2 km s$^{-1}$.

We used the chopper wheel method.
We then estimated the absolute intensity calibration error at 10\%, which is a typical value for the chopper wheel method.

We employed the Smoothed Bandpass Calibration (SBC) method \citep{2012PASJ...64..118Y}.
The SBC method allows us to reduce the time for observing off-source positions.
The scan pattern was set as 20 seconds and 5 seconds for on-source and off-source positions, respectively.
We applied 60 channel-smoothing only for off-source spectra.

\section{Results and Analysis}

\subsection{Results} \label{sec:res}

The rotational line of HC$_{5}$$^{15}$N was clearly detected with the signal-to-noise ratio of 7, as shown in Figure \ref{fig:f1}.
The on-source integration time is 45 hours 2 minutes, and the rms noise level in the line-free region is 2.4 mK in $T_{\rm {A}}^{*}$ with the velocity resolution of 0.2 km s$^{-1}$.
We fitted the spectra with a Gaussian profile, and obtained the spectral line parameters.
The value of $V_{\rm {LSR}}$ of the line is $5.7 \pm 0.3$ km s$^{-1}$, which is consistent with the systemic velocity of TMC-1 CP (5.85 km s$^{-1}$).
The peak intensity ($T_{\rm {A}}^{*}$), the line width (FWHM), and the integrated intensity ($\int T^{\ast}_{\mathrm A}dv$) are $17 \pm 2$ mK, $0.42 \pm 0.07$ km s$^{-1}$, and $0.007 \pm 0.002$ K km s$^{-1}$ {\bf {($1\sigma$)}}, respectively.
The errors of the line parameters were derived from the Gaussian fitting.

We verified the line identification from the line width and rest frequency.
First, the derived line width ($0.42 \pm 0.07$ km s$^{-1}$) is consistent with the typical value in TMC-1 CP (0.5 km s$^{-1}$, \citet{2004PASJ...56...69K}) and the spectrum does not appear a spiky instrumental spurious.
Second, there is no other detectable line in TMC-1 CP in the 23.35$-$23.4 GHz band according to the Splatalogue database for astronomical spectroscopy\footnote{http://www.cv.nrao.edu/php/splat/}.
Thus, we concluded that the detected emission line should be identified as HC$_{5}$$^{15}$N.

\begin{figure}[!h]
 \begin{center}
  \includegraphics[width=7.5cm]{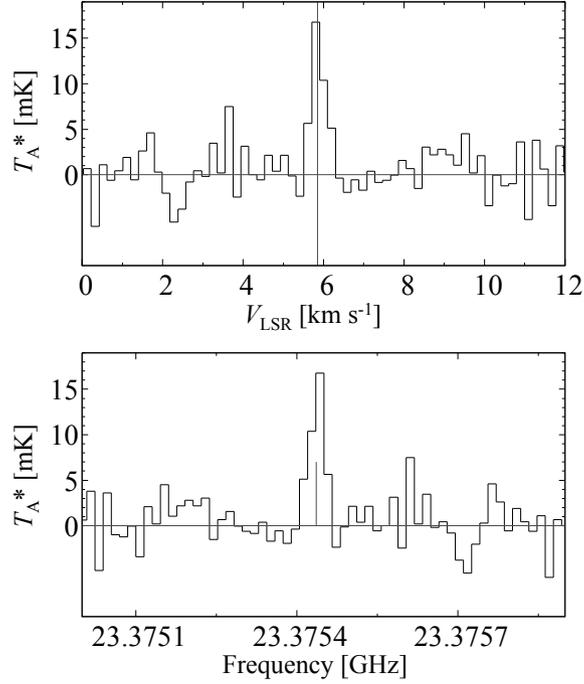} \label{fig:f1}
 \end{center}
\caption{Spectra of HC$_{5}$$^{15}$N ($J=9-8$) in TMC-1 CP. Upper; The horizontal axis is the velocity unit. The vertical line shows the systemic velocity of TMC-1 CP (5.85 km s$^{-1}$). Lower; The horizontal axis is the frequency unit. The vertical line indicates the rest frequency of the line (23.37544 GHz, \citet{2005JChPh.122s4319S})}\label{fig:f1}
\end{figure}


\subsection{Analysis} \label{sec:ana}

We calculated the column density assuming the local thermodynamic equilibrium (LTE).
We used the following formulae \citep{2016ApJ...817..147T}:

\begin{equation} \label{tau}
\tau = - {\mathrm {ln}} \left[1- \frac{T^{\ast}_{\rm A} }{f\eta_{\rm B} \left\{J(T_{\rm {ex}}) - J(T_{\rm {bg}}) \right\}} \right],  
\end{equation}
where
\begin{equation} \label{tem}
J(T) = \frac{h\nu}{k}\Bigl\{\exp\Bigl(\frac{h\nu}{kT}\Bigr) -1\Bigr\} ^{-1},
\end{equation}  
and
\begin{eqnarray} \label{col}
    N = \tau \frac{3h\Delta v}{8\pi ^3}\sqrt{\frac{\pi}{4\mathrm {ln}2}}Q\frac{1}{\mu ^2}\frac{1}{J_{\rm {lower}}+1}\exp\Bigl(\frac{E_{\rm {lower}}}{kT_{\rm {ex}}}\Bigr)
                    \times \Bigl\{1-\exp\Bigl(-\frac{h\nu }{kT_{\rm {ex}}}\Bigr)\Bigr\} ^{-1}.
\end{eqnarray} 
In Equation (\ref{tau}), $T^{\ast}_{\rm A}$ denotes the antenna temperature, {\it f} the beam filling factor, $\eta_{\rm B}$ the main beam efficiency (0.8, Section \ref{sec:obs}), and $\tau$ the optical depth.
Since a size of the emitting region of carbon-chain molecules in TMC-1 CP is approximately 2.5' according to the mapping observations by \citet{1992ApJ...394..539H}, we used 0.8 for {\it f}. 
$T_{\rm {ex}}$ is the excitation temperature, $T_{\rm {bg}}$ is the cosmic microwave background temperature ($\approx 2.7$ K), and {\it J}({\it T}) in Equation (\ref{tem}) is the Planck function. 
In the calculation, we adopted the excitation temperature of 6.5 K of the normal species \citep{2016ApJ...817..147T}.
In Equation (\ref{col}), {\it N} denotes the column density, $\Delta v$ the line width (FWHM), $Q$ the partition function, $\mu$ the permanent electric dipole moment of HC$_{5}$$^{15}$N (4.33 D; \cite{1976JMoSp..62..175A}), and $E_{\rm {lower}}$ the energy of the lower rotational energy level. 

The derived column density of HC$_{5}$$^{15}$N is ($1.9 \pm 0.5$)$\times 10^{11}$ cm$^{-2}$ ($1\sigma$).
The $^{14}$N/$^{15}$N ratio of HC$_{5}$N is determined to be $323 \pm 80$ ($1\sigma$), using the column density of the normal species (($6.2 \pm 0.3) \times 10^{13}$ cm$^{-2}$, \cite{2016ApJ...817..147T}).
\citet{2016ApJ...817..147T} derived the optical depth of the normal species to be $1.084 \pm 0.014$ ($1\sigma$), and the uncertainty of the column density of the normal species should be small.
The errors contain the $1\sigma$ errors from the Gaussian fitting and the 10\% absolute intensity uncertainty from the chopper wheel method.


\section{Discussion}

\subsection{Deriving the $^{14}$N/$^{15}$N ratio of HC$_{5}$N using the double isotope method}

The $^{14}$N/$^{15}$N ratios are often derived with the double isotope method, using the two isotopologues (e.g., \cite{2013Icarus...223...582}).
In the case of HC$_{5}$N, we can calculate five cases independently, and reduce the uncertainty in the $^{14}$N/$^{15}$N ratio statistically.
We observed the normal species, the five $^{13}$C isotopologues, and the $^{15}$N isotopologue simultaneously.
Therefore, we can exclude the pointing and calibration errors.
We calculated the $^{14}$N/$^{15}$N ratios of HC$_{5}$N with the double isotope method, using the following formula: 

\begin{equation} \label{dou}
\frac{{\mathrm {HC}}_{5}\,^{14}{\mathrm N}}{{\mathrm {HC}}_{5}\,^{15}{\mathrm N}} = \Bigl< \frac{{\mathrm {HC}}_{m}\,^{13}{\mathrm {CC}}_{n}{\mathrm N}}{{\mathrm {HC}}_{5}\,^{15}{\mathrm N}}\times \frac{{\mathrm {HC}}_{5}\,^{14}{\mathrm N}}{{\mathrm {HC}}_{m}\,^{13}{\mathrm {CC}}_{n}{\mathrm N}} \Bigr>
\end{equation}
where $m, n = 0-4$ ($m+n = 4$). 
In Equation (\ref{dou}), HC$_{5}$$^{14}$N/HC$_{m}$$^{13}$CC$_{n}$N represents the $^{12}$C/$^{13}$C ratios (column of $^{12}$C/$^{13}$C in Table \ref{tab:t1}, taken from \citet{2016ApJ...817..147T}).
We derived HC$_{m}$$^{13}$CC$_{n}$N/HC$_{5}$$^{15}$N using the integrated intensities (column of x/HC$_{5}$$^{15}$N in Table \ref{tab:t1}), because both $^{15}$N and $^{13}$C isotopologues are optically thin.
< > denotes the mean value.
We calculated the HC$_{5}$$^{14}$N/HC$_{5}$$^{15}$N ratio in the case of $m = 0$ and $n=4$ using Equation (\ref{dou}).
We repeated the calculations until $m=4$ and $n=0$, and finally we averaged the five HC$_{5}$$^{14}$N/HC$_{5}$$^{15}$N ratios.
The calculated five $^{14}$N/$^{15}$N ratios of HC$_{5}$N are summarized in Table \ref{tab:t1} (column of $^{14}$N/$^{15}$N), and these values are consistent with each other.
The averaged $^{14}$N/$^{15}$N ratio of HC$_{5}$N was derived to be $344 \pm 53$ (1$\sigma$).
The derived $^{14}$N/$^{15}$N ratio of HC$_{5}$N using the double isotope method is consistent with $323 \pm 80$ ($1\sigma$) derived in Section \ref{sec:ana}.
The $^{14}$N/$^{15}$N ratio of HC$_{5}$N in TMC-1 CP is smaller than that of the elemental ratio of $\sim 440$ \citep{2011Sci...332.1533M}.

\begin{table} 
  \tbl{The integrated intensity ratios between the five $^{13}$C isotopologues and $^{15}$N isotopologue, $^{12}$C/$^{13}$C ratios, and $^{14}$N/$^{15}$N ratios of HC$_{5}$N}{%
  \begin{tabular}{lccc}
     \hline
     $^{13}$C isotopologues (X) & X/HC$_{5}$$^{15}$N & $^{12}$C/$^{13}$C & $^{14}$N/$^{15}$N \\ 
     \hline
     H$^{13}$CCCCCN & 3.6 (0.5) & 98 (14) & 352 (98) \\      
     HC$^{13}$CCCCN & 3.4 (0.6) & 101 (14) & 345 (96) \\ 
     HCC$^{13}$CCCN & 3.5 (0.5) & 95 (12) & 331 (90) \\
     HCCC$^{13}$CCN & 3.6 (0.5) & 93 (13) & 337 (94) \\
     HCCCC$^{13}$CN & 4.2 (0.6) & 85 (11) & 355 (96) \\
     Average value        &  &  & 344 (53) \\
     \hline
   \end{tabular}}\label{tab:t1}
\begin{tabnote}
The numbers in parenthesis represent one standard deviation.
The integrated intensities of $^{13}$C isotopologues of HC$_{5}$N and the $^{12}$C/$^{13}$C ratios were derived by \citet{2016ApJ...817..147T}.
\end{tabnote}
\end{table}

\subsection{Comparison of the $^{14}$N/$^{15}$N ratios}

We also derived the $^{14}$N/$^{15}$N ratio of HC$_{3}$N in TMC-1 CP using the double isotope method from the previous observational results with the Nobeyama 45-m telescope.
We derived the integrated intensity ratios (column of x/HC$_{3}$$^{15}$N in Table \ref{tab:t2}) using the $J=4-3$ rotational transition at the 36 GHz band from \citet{2004PASJ...56...69K}.
\citet{1998A&A...329.1156T} observed the normal species and the three $^{13}$C isotopologues of HC$_{3}$N, and derived the three $^{12}$C/$^{13}$C ratios.
We assumed that the error in the result of \citet{2004PASJ...56...69K} is 20\%.
The derived three $^{14}$N/$^{15}$N ratios of HC$_{3}$N are listed in Table \ref{tab:t2} (column of $^{14}$N/$^{15}$N).
We estimated its $^{14}$N/$^{15}$N ratio to be $257 \pm 54$ ($1\sigma$).

We derived the column density of HC$_{3}$$^{15}$N using Equations (\ref{tau}) - (\ref{col}) and the line parameters of its $J=4-3$ rotational line taken from \citet{2004PASJ...56...69K}.
We used the line, because we can assume that the filling factor is almost unity.
We then used the beam filling factor of unity and the main beam efficiency of 0.8\footnote{http://www.nro.nao.ac.jp/~nro45mrt/html/prop/eff/eff\_before2001.html\#period5}.
We used the excitation temperature of its normal species (7.1 K, \cite{1998A&A...329.1156T}).
We assumed that the uncertainty of the integrated intensity is $20\%$.
The derived column density of HC$_{3}$$^{15}$N is ($5.9 \pm 0.5$)$\times 10^{11}$ cm$^{-2}$.
The column density of the normal species is ($1.6 \pm 0.1$)$\times 10^{14}$ cm$^{-2}$ \citep{1998A&A...329.1156T}.
Therefore, the derived $^{14}$N/$^{15}$N ratio of HC$_{3}$N is $270 \pm 57$ ($1\sigma$)\footnote{We divided the integrated intensity of HC$_{3}$$^{15}$N by a scaling factor of 1.3 to correct the difference between \citet{1998A&A...329.1156T} and \citet{2004PASJ...56...69K}. The scaling factor of 1.3 was derived by comparison of the integrated intensities of H$^{13}$CCCN ($J=4-3$) between \citet{1998A&A...329.1156T} and \citet{2004PASJ...56...69K}.}.
The $^{14}$N/$^{15}$N ratios derived by the two methods are well consistent with each other.

\begin{table} 
  \tbl{The integrated intensity ratios between the five $^{13}$C isotopologues and $^{15}$N isotopologue, $^{12}$C/$^{13}$C ratios, and $^{14}$N/$^{15}$N ratios of HC$_{3}$N}{%
  \begin{tabular}{lccc}
     \hline
     $^{13}$C isotopologues (X) & X/HC$_{3}$$^{15}$N & $^{12}$C/$^{13}$C & $^{14}$N/$^{15}$N \\ 
     \hline
     H$^{13}$CCCN & 3.2 (0.4) & 79 (11) & 253 (62) \\      
     HC$^{13}$CCN & 3.5 (0.5) & 75 (10) & 265 (64) \\ 
     HCC$^{13}$CN & 4.6 (0.7) & 55 (7) & 254 (60) \\
     Average value &  &  & 257 (54) \\
      \hline
   \end{tabular}}\label{tab:t2}
\begin{tabnote}
The numbers in parenthesis represent one standard deviation.
The integrated intensity ratios between the $^{13}$C and $^{15}$N isotopologues were derived using the results of \citet{2004PASJ...56...69K}.
The $^{12}$C/$^{13}$C ratios were derived by \citet{1998A&A...329.1156T}.
The $^{14}$N/$^{15}$N ratios were derived in this paper.
\end{tabnote}
\end{table}

\subsection{Main formation mechanism of HC$_{5}$N}
The elemental $^{14}$N/$^{15}$N ratio in the local interstellar medium was estimated to be $441 \pm 6$ from the solar wind \citep{2011Sci...332.1533M}.
In addition, although the $^{14}$N/$^{15}$N ratio of CN could not be derived due to a non-thermal intensity ratio of the hyperfine lines of the normal species of CN in TMC-1 \citep{2002ApJ...578..211S}, $^{15}$N generally tends to concentrate in CN molecules in cold environments \citep{2008ApJ...689.1448R}.
The $^{14}$N/$^{15}$N ratio of HC$_{3}$N in TMC-1 CP is smaller than that of the elemental ratio of $441 \pm 6$ \citep{2011Sci...332.1533M}. 
The results suggest that N in HC$_{3}$N does not come from nitrogen atoms, but originates from CN, which agrees with the HC$_{3}$N formation pathway suggested from the $^{13}$C isotopic fractionation \citep{1998A&A...329.1156T}.
In addition, the small $^{14}$N/$^{15}$N ratio of HC$_{3}$N implies that $^{15}$N is concentrated in CN molecules in TMC-1 CP, as suggested by the model calculation \citep{2008ApJ...689.1448R}.
On the other hand, the derived $^{14}$N/$^{15}$N ratio of HC$_{5}$N in TMC-1 CP is slightly higher than that of HC$_{3}$N in TMC-1 CP, if we take the ratio determined from the double isotope method. 
Hence, the results may suggest that the main formation mechanism of HC$_{5}$N is different from that of HC$_{3}$N.

The possible formation pathways of HC$_{5}$N in TMC-1 CP were discussed in \citet{2016ApJ...817..147T}, and they categorized the pathways into three mechanisms as following:\\
\noindent Mechanism 1: the reactions of C$_{4}$H$_{2}$ + CN,\\
\noindent Mechanism 2: the growth of the cyanopolyyne carbon chains via C$_{2}$H$_{2}$$^{+}$ + HC$_{3}$N, and \\
\noindent Mechanism 3: the reactions between hydrocarbon ions and nitrogen atoms followed by electron recombination reactions. \\
Only Mechanism 3 does not contain CN molecules intrinsically, and N in HC$_{5}$N originates from nitrogen atoms.
The derived $^{14}$N/$^{15}$N ratio of HC$_{5}$N in TMC-1 CP ($344 \pm 53$ (1$\sigma$)) suggests that Mechanism 3 dominates the formation of HC$_{5}$N.
The small difference in the $^{14}$N/$^{15}$N ratio between HC$_{5}$N and the elemental ratio seems to imply that the reactions containing CN molecules (Mechanisms 1 and 2) partly contribute to the formation of HC$_{5}$N, as suggested by \citet{2016ApJ...817..147T}.

\section{Conclusions}

We have detected HC$_{5}$$^{15}$N from TMC-1 CP for the first time in the interstellar medium.
Its column density is derived to be ($1.9 \pm 0.5$)$\times 10^{11}$ cm$^{-2}$ ($1\sigma$).
The $^{14}$N/$^{15}$N ratio of HC$_{5}$N is calculated to be $323 \pm 80$ ($1\sigma$).
In addition, we evaluate its $^{14}$N/$^{15}$N ratio using the double isotope method, and the value is determined at $344 \pm 53$ ($1\sigma$).
The $^{14}$N/$^{15}$N ratios derived from the column density and the double isotope method are consistent with each other.
We also derived the $^{14}$N/$^{15}$N ratio of HC$_{3}$N in TMC-1 CP using the previous observational results.
The ratio is derived to be $257 \pm 54$ (1$\sigma$) from the double isotope method.
The $^{14}$N/$^{15}$N ratio of HC$_{3}$N is derived to be $270 \pm 57$ (1$\sigma$) using the column density.
The double isotope method resulted in slightly higher $^{14}$N/$^{15}$N ratio of HC$_{5}$N than HC$_{3}$N in TMC-1 CP.
The results may support our previous work, proposing that the main formation mechanism of HC$_{5}$N is the ion-molecule reactions between hydrocarbon ions and nitrogen atoms followed by the electron recombination reactions, but other formation pathways including CN partly contribute to the formation of HC$_{5}$N.

\begin{ack}
We deeply appreciate the anonymous referee for suggestions, which are very constructive to make our discussion.
We are grateful to the staff of the Nobeyama Radio Observatory.
Nobeyama Radio Observatory is a branch of the National Astronomical Observatory of Japan, National Institutes of Natural Sciences.
\end{ack}



\begin{thebibliography}{}
\bibitem[Alexander et al.(1976)]{1976JMoSp..62..175A} Alexander, A. J., Kroto, H. W., \& Walton, D. R. M.\ 1976, Journal of Molecular Spectroscopy, 62, 175
\bibitem[Hily-Blant et al. (2013)]{2013Icarus...223...582} Hily-Blant, P., Bonal, L., Faure, A., \& Quirico, E.\ 2013, Icarus, 223, 582
\bibitem[Hirahara et al.(1992)]{1992ApJ...394..539H} Hirahara, Y., Suzuki, H., Yamamoto, S., et al.\ 1992, \apj, 394, 539
\bibitem[Hirota et al.(2009)]{2009ApJ...699..585H} Hirota, T., Ohishi, M., \& Yamamoto, S.\ 2009, \apj, 699, 585
\bibitem[Jaber Al-Edhari et al.(2017)]{2017A&A...597A..40J} Jaber Al-Edhari, A., Ceccarelli, C., Kahane, C., et al.\ 2017, \aap, 597, A40
\bibitem[Kaifu et al.(2004)]{2004PASJ...56...69K} Kaifu, N., Ohishi, M., Kawaguchi, K., et al.\ 2004, \pasj, 56, 69
\bibitem[Kaiser et al.(1991)]{1991ApJ...379..267K} Kaiser, M. E., Hawkins, I., \& Wright, E. L.\ 1991, \apj, 379, 267
\bibitem[Marty et al.(2011)]{2011Sci...332.1533M} Marty, B., Chaussidon, M., Wiens, R. C., Jurewicz, A. J. G., \& Burnett, D. S.\ 2011, Science, 332, 1533
\bibitem[Rodgers \& Charnley(2008)]{2008ApJ...689.1448R} Rodgers, S. D., \& Charnley, S. B.\ 2008, \apj, 689, 1448
\bibitem[Sakai et al.(2007)]{2007ApJ...663...1174} Sakai, N., Ikeda, M., Morita, M., Sakai, T., Takano, S., Osamura, Y., \& Yamamoto, S.\ 2007, \apj, 663, 1174
\bibitem[Sakai et al.(2010)]{2010A&A...512A...31S} Sakai, N., Saruwatari, O., Sakai, T., Takano, S., \& Yamamoto, S.  2010, \aap, 512, A31
\bibitem[Sakai \& Yamamoto(2013)]{2013ChRv...113...8981} Sakai, N., \& Yamamoto, S.\  2013, ChRv, 113, 8981
\bibitem[Sakai et al.(2013)]{2013JPCA...117...9831} Sakai, N., Takano, S., Sakai, T., Shiba, S., Sumiyoshi, Y., Endo, Y., \& Yamamoto, S.\  2013, JPCA, 117, 9831
\bibitem[Sanz et al.(2005)]{2005JChPh.122s4319S} Sanz, M. E., McCarthy, M. C., \& Thaddeus, P.\ 2005, \jcp, 122, 194319
\bibitem[Savage et al.(2002)]{2002ApJ...578..211S} Savage, C., Apponi, A. J., Ziurys, L. M., \& Wyckoff, S.\ 2002, \apj, 578, 211
\bibitem[Suzuki et al.(1992)]{1992ApJ...392..551S} Suzuki, H., Yamamoto, S., Ohishi, M., et al.\ 1992, \apj, 392, 551
\bibitem[Takano et al.(1998)]{1998A&A...329.1156T} Takano, S., Masuda, A., Hirahara, Y., et al.\ 1998, \aap, 329, 1156
\bibitem[Taniguchi et al.(2016a)]{2016ApJ...817..147T} Taniguchi, K., Ozeki, H., Saito, M., Sakai, N., Nakamura, F., Kameno, S., Takano, S., \& Yamamoto, S\ 2016a, \apj, 817, 147
\bibitem[Taniguchi et al.(2016b)]{2016ApJ...830..106T} Taniguchi, K., Saito, M., \& Ozeki, H.\ 2016b, \apj, 830, 106
\bibitem[Tatematsu et al.(2017)]{2017ApJS..228...12T} Tatematsu, K., Liu, T., Ohashi, S., et al.\ 2017, \apjs, 228, 12
\bibitem[Yamaki et al.(2012)]{2012PASJ...64..118Y} Yamaki, H., Kameno, S., Beppu, H., Mizuno, I., \& Imai, H.\ 2012, \pasj, 64, 118
\bibitem[Yoshida et al.(2015)]{2015ApJ...807...66Y} Yoshida, K., Sakai, N., Tokudome, T., et al.\ 2015, \apj, 807, 66

\end{thebibliography}
\end{document}